\documentclass[epj]{svjour}
\usepackage{graphics}
\begin{document}
\title{Networks of equities in financial markets}
\author{G. Bonanno\inst{1,2}, G. Caldarelli\inst{1}, 
F. Lillo\inst{2,3}, S. Miccich\`e\inst{2,3}, 
N. Vandewalle\inst{4}, \and R. N. Mantegna\inst{2,3}% etc
%\thanks{\emph{Permanent address:} Dipartimento di Fisica e Tecnologie Relative, 
%Universit\`a  di Palermo, Viale delle Scienze, I-90128 Palermo, Italia}%
}                     % Do not remove
\institute{Istituto Nazionale per la Fisica della Materia,
Unit\`a di Roma, Roma "La Sapienza", Roma, Italy 
  \and 
Dipartimento di Fisica e Tecnologie Relative, Universit\`a
di Palermo, Viale delle Scienze, I-90128, Palermo, Italy
\and
Istituto Nazionale per la Fisica della Materia,
Unit\`a di Palermo, Palermo, Italy 
  \and 
GRASP, Institut de Physique B5, Universit\'e de Li\`ege, B-4000 
Li\`ege, Belgium  
}
\date{Received: date / Revised version: date}

\abstract{We review the recent approach of correlation based networks of financial
equities. We investigate portfolio of stocks at different time horizons, financial
indices and volatility time series and we show that meaningful economic information
can be extracted from noise dressed correlation matrices. We show that the method 
can be used to falsify widespread market models by directly comparing the topological 
properties of networks of real and artificial markets.
 % end of PACS codes
} %end of abstract

\authorrunning{Bonanno {\it et al.}}
\titlerunning{Networks of equities in financial markets}
\maketitle
\section{Introduction}
\label{intro}

The study of topological properties of networks has recently received a lot of
attention. In particular, it has been shown that many natural  and social systems
display unexpected statistical properties  of links connecting different elements of
the system \cite{Watts98,Barabasi99} and cannot therefore be described in terms of
random graphs \cite{Erdos}. The topological properties of several graphs describing
physical  and social systems have been recently investigated. Examples are the World
Wide Web \cite{Albert99}, Internet \cite{Caldarelli2000,Pastor2001}, and social
networks \cite{Newman2002}.  In the networks investigated in these papers (and in many
others) the links  represent relation between nodes which are either present or absent
in a given instant of time.  By contrast we have recently started the investigation of
correlation based networks, i.e. networks used to visualize the structure of pair
cross correlations among a set of time series. From a set of $n$ time series one can
extract the correlation coefficient between any pair of variables. If we identify the
different time series with the nodes of the network, each pair of nodes can be thought
to be connected by an arc with a weight related to the correlation coefficient between
the two time series. The network is therefore completely connected.  By introducing a
suitable filtration of the network one can remove the less relevant information by
removing the weakest links. In fact, it is known that the finiteness of time series
can introduce spurious correlation. In principle there are many different ways of
filtering the correlation matrix in order to obtain noise filtered information.  In
this context we have focused mainly on financial markets
\cite{Mantegna99,Vandewalle2001} and on a particular type of network that can be
obtained form the correlation matrix, specifically the minimum spanning tree. Spanning
trees are particular types of graphs that connect all the vertices in a graph without
forming any loop.

The presence of a high degree of cross-correlation between the synchronous time
evolution of a set of equity returns is a well known empirical fact observed in
financial markets \cite{Markowitz59,Elton95,Campbell97}. For a time horizon of one
trading day correlation coefficient as high as 0.7 can be observed for some pair of
equity returns belonging to the same economic sector.

The study of cross-correlation of a set of financial equities has also practical
importance since it can improve the ability to model composed financial entities such
as, for example, stock portfolios. There are different approaches to address this
problem. The most common one is the principal component analysis of the correlation
matrix of the data \cite{Elton71}. Recently an investigation of the properties of the
correlation matrix has been performed by physicists by using the perspective and
theoretical results of the random matrix theory \cite{Laloux99,Plerou99}.  As
mentioned above, another approach is the correlation based clustering analysis which
allows to obtain clusters of stocks starting from the time series of price returns.
Different algorithms exist to perform cluster analysis in finance 
\cite{Mantegna99,Panton76,Kullmann2000,Bernaschi2000,Giada2001,Marsili2002}.

In previous work, some of us have shown that a specific correlation based
clustering method gives a meaningful taxonomy for stock return time series \cite{Mantegna99,Bonanno2001,Bonanno2003}, for market index returns 
of worldwide stock exchanges \cite{Bonanno2000} and for volatility increments of stock return time series \cite{Micciche2003}. Here we review the results obtained in these previous studies 
and discuss them from a unified perspective. Specifically, 
Sect. 2 discusses the correlation based clustering method, Sect. 3
focuses on the properties of networks detected in a portfolio
of stocks when stock returns are sampled at different time horizons.
Sect. 4 discusses the properties of networks observed by investigating
stock indices of stock exchanges located all over the world
and Sect. 5 discusses the case of financial networks obtained starting 
from volatility time series. Sect. 6 is about the comparison 
of topological properties of real data with the ones of simple 
and widespread models of market activity. Finally, in Sect. 7 we
draw our conclusions.

\section{A financial network obtained by a correlation-based 
filtering procedure}

In Ref. \cite{Mantegna99}, it has been proposed 
a correlation based method able to detect economic information
present in a correlation coefficient matrix. This method is 
a filtering procedure based on
the estimation of the subdominant ultrametric \cite{Rammal86} associated with a
metric distance obtained form the correlation coefficient matrix of set of $n$
stocks. This procedure,  already used in other fields, allows 
to obtain a metric
distance and to extract from it a minimum spanning tree (MST) and 
a hierarchical tree from a
correlation coefficient matrix by means of a well defined algorithm known as
nearest neighbor single linkage clustering algorithm \cite{Mardia79}. 
This allows to
reveal geometrical  (throughout the MST) and taxonomic (throughout the
hierarchical tree) aspects of the correlation present among stocks.

The network  
is obtained by filtering the relevant information present
in the correlation coefficient matrix of the original time series
of stock returns. This is done (i) by  determining the synchronous 
correlation coefficient of  the
difference of logarithm of stock price computed at  a selected time horizon, 
(ii) by calculating a metric distance  between all the pair of stocks and (iii)
by selecting the  subdominant ultrametric distance associated to the 
considered metric distance. The subdominant ultrametric is the ultrametric
structure closest to the original metric structure \cite{Rammal86}.

The correlation coefficient is defined as 
\begin{equation}
\rho_{ij} (\Delta t) \equiv \frac{\langle r_i r_j\rangle -\langle r_i\rangle\langle r_j\rangle}
{\sqrt{(\langle r_i^2\rangle-\langle r_i\rangle^2)(\langle r_j^2\rangle -\langle r_j\rangle^2)}}
\end{equation}
where $i$ and $j$ are  numerical labels of the stocks, 
$r_i=\ln P_i(t)-\ln P_i(t-\Delta t)$, $P_i(t)$ is the value of 
the stock price $i$ at the trading time $t$ and $\Delta t$ is the
time horizon which is, in the present Section, one trading day. 
The correlation coefficient 
for logarithm price differences (which almost coincides with 
stock returns) is 
computed between all the possible pairs of stocks present in 
the considered portfolio. 
The empirical statistical average, indicated in this paper 
with the symbol $\langle.\rangle$,
is here a temporal average always
performed over the investigated time period. 

By definition, $\rho_{ij} (\Delta t)$ can vary from -1 (completely 
anti-correlated pair of stocks) to 1 (completely correlated 
pair of stocks).
When $\rho_{ij} (\Delta t)=0$ the two stocks are uncorrelated. 
The matrix of correlation coefficient is a symmetric matrix with 
$\rho_{ii}(\Delta t)=1$ in the main diagonal. 
Hence for each value of $\Delta t$, $n(n-1)/2$ correlation 
coefficients characterize 
each correlation coefficient matrix completely.

A metric distance between pair of stocks can be rigorously 
determined \cite{Gower66} by defining
\begin{equation}
d_{i,j} (\Delta t)=\sqrt{2(1-\rho_{ij}(\Delta t))} .
\end{equation}
With this choice $d_{i,j}(\Delta t)$ fulfills the three axioms of a metric --
(i) $d_{i,j}(\Delta t)=0$ if and only if $i=j$; 
(ii) $d_{i,j}(\Delta t)=d_{j,i} (\Delta t)$ and (iii)
$d_{i,j} (\Delta t) \le d_{i,k} (\Delta t) +d_{k,j} (\Delta t)$. 
The distance matrix ${\bf{D}} (\Delta t)$ is 
then used to determine the MST connecting the $n$ stocks. 

The MST, a theoretical concept of graph theory 
\cite{West96}, is the spanning tree of shortest length.
A spanning tree is a graph without loops connecting 
all the $n$ nodes with $n-1$ links. We have seen that
the original fully connected graph is metric with 
distance $d_{i,j}$ which is decreasing with $\rho_{ij}$.
Therefore the MST selects the $n-1$
stronger (i.e. shorter) links which span all the nodes. 
The MST allows to obtain, 
in a direct and essentially unique way, the subdominant 
ultrametric distance matrix ${\bf{D}}^<(\Delta t)$ 
and the hierarchical organization of the elements 
(stocks in our case) of the investigated data set.

The subdominant ultrametric distance between objects $i$ and $j$, 
i.e. the element $d^<_{i,j}$ of the ${\bf{D}}^<(\Delta t)$  
matrix, is the maximum value 
of the metric distance $d_{k,l}$ detected by moving 
in single steps from $i$ to $j$ through the path connecting 
$i$ and $j$ in the MST. 
The method of constructing a MST linking a set of $n$ objects 
is direct and it is known in multivariate analysis as the
nearest neighbor single linkage cluster analysis \cite{Mardia79}. 
A pedagogical exposition of the determination of the
MST in the contest of financial time series is provided in ref. 
\cite{MS2000}. Subdominant ultrametric space \cite{Rammal86} has been 
fruitfully used in the description of frustrated complex systems. 
The archetype of this kind of systems is a spin glass \cite{Mezard87}. 

As an example of the results obtained with this method
here we briefly discuss the results obtained in 
ref. \cite{Bonanno2001}, by investigating a set of 100 highly 
capitalized stocks traded in 
the major US equity markets during the period January 1995 - 
December 1998. At that time, most of them were  used to compute the
Standard and Poor's 100 index. The prices are transaction prices 
stored in the {\it Trade and Quote} database of the 
New York Stock Exchange. 

The time horizons investigated in the cited study 
varies from $\Delta t=d=6$ h and $30$ min (a trading day time interval), 
to $\Delta t=d/20=19$ min and $30$ sec.

\begin{figure}
\resizebox{1.0\columnwidth}{!}{\includegraphics{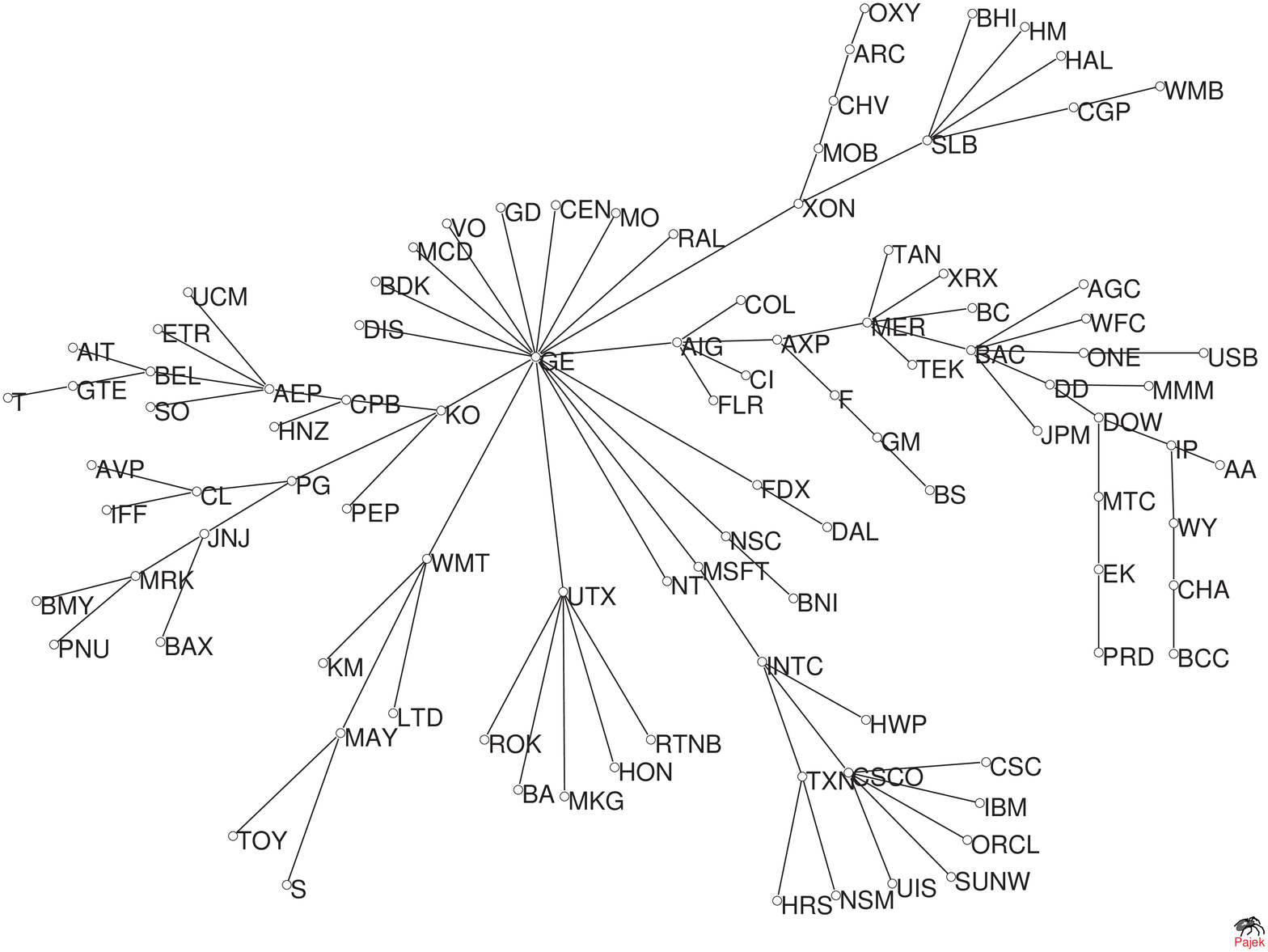}}
\caption{Minimum spanning tree of 100 highly capitalized 
stocks traded in the US equity markets. The filtering
procedure has been obtained by considering the correlation
coefficient of stock returns time series computed at  
a 1 trading day time horizon (6 h and 30 min). Each circle 
represents a stock labeled by its tick symbol. The minimum 
spanning tree presents a large amount of stocks having a single link
and some stocks having several links. Some of these stocks act
as a ``hub" of a local cluster. Examples are INTC and CSCO for 
technology stocks, AIG, BAC and MER for financial stocks
and AEP for utilities stocks. The stock GE (General Electric Co.) 
links a relatively large number of stocks belonging to various 
sectors.}
\label{figA}       % Give a unique label
\end{figure}

In Fig.~\ref{figA} we show the minimal spanning tree obtained in this
investigation with a time horizon equal to one trading day.
Stocks are identified with their tick symbols. 
Information about the 
company indicated by each tick symbol can be easily find
in several financial web pages such as, for example, 
http://www.quicken.com .
Cluster of stocks which are homogeneous with respect to the 
economic sectors of firms are clearly observed.
Prominent examples of clusters are the ones of (i) oil companies
which is, to be precise, a cluster composed by two separated 
sub-clusters, one including the companies SLB, HAL, BHI,
CGP and WMB 
(companies which are providing financial services to 
the oil industry and companies of the gas industry) and the other one including
MOB, CHV, XON, ARC, OXY (companies of the oil industry); (ii) 
financial (JPM, BAC, MER, USB, ONE, WFC, APX, etc)
and consumer/non-cyclical companies (KO, GE, PG, CL, AVP, JNJ, etc);
(iii) technology companies (MSFT, INTC, TXN, CSCO, NSM, IBM, HWP,
ORCL); (iv) basic materials companies (AA, WY, CHA, IP, BCC),
and (v) utility companies (BEL, AIT, GTE, SO, AEP, UCM, ETR).

Equity time series are then carrying economic information which 
can be detected by using specialized filtering procedures. 
Therefore, price time series in a financial market reflect information about 
the economic sector of activity of the company.  This information is usually 
dressed by the noise due to statistical fluctuations. Filtering procedures,
like the one we are proposing, are able to undress the signals from the noise
and reveal the more relevant information.

\section{Minimal spanning trees of stock portfolios at 
different time horizons}

\begin{figure*}
\resizebox{1.8\columnwidth}{!}{\includegraphics{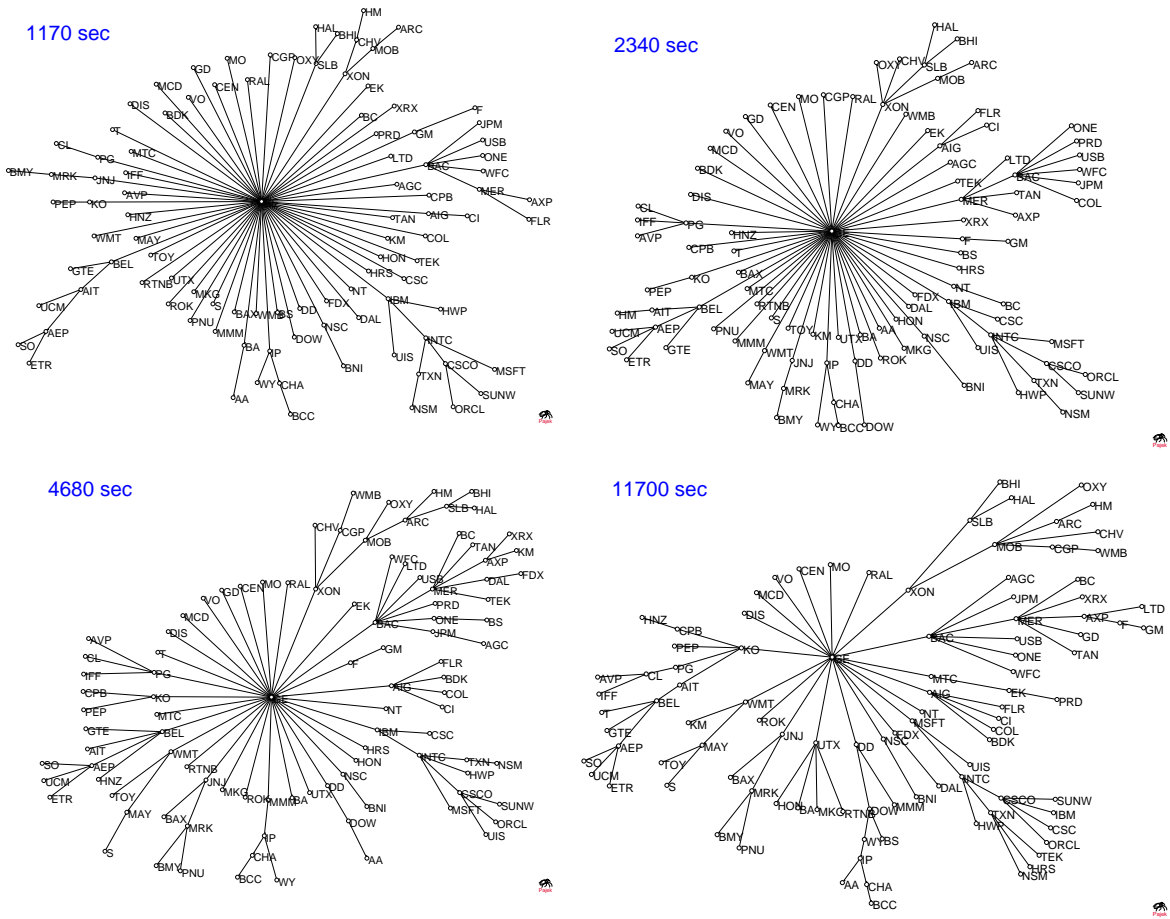}}
\vspace*{0.5cm}       % Give the correct figure height in cm
\caption{Minimum spanning tree of 100 highly capitalized 
stocks traded in the US equity markets. The time scale $\Delta t$ used to 
compute the correlation coefficients between stocks is smaller 
than one trading day. Specifically we show the MST obtained with
$\Delta t= d/20$ (top left), $\Delta t= d/10$ (top right),
$\Delta t= d/5$ (bottom left), and $\Delta t= d/2$ (bottom right), where $d$ is
one trading day.}
\label{figB}       % Give a unique label
\end{figure*}

In this section we discuss how the correlation structure of a portfolio
of stocks changes when the time horizon used to compute the
correlation coefficient is progressively decreased to an intraday  time scale.
It is known since 1979 that the degree of cross-correlation 
diminishes by diminishing the time horizon used to compute 
stock returns \cite{Epps79}. This phenomenon is sometime 
addressed as ``Epps effect". The existence of this phenomenon 
motivates us to investigate the nature and the properties 
of the network associated to a given financial portfolio
as a function of the time horizon used to record stock
return time series.
 
In Ref. \cite{Bonanno2001}, some of us used the high-frequency 
data of the transactions occurring in the US equity markets which 
are recorded in the {\it Trade and Quote}
database of the New York Stock Exchange. By using this database we
are able to investigate comovements of a set of highly capitalized 
stocks for daily and intra daily time horizons.

A clear modification of the hierarchical organization 
of the set of stocks investigated is detected when one changes 
the time horizon used to determine stock returns.
The structure of the considered set of 100 US stocks 
changes its nature moving from a
complex organization to a progressively elementary one 
when the time horizon 
of price changes varies from $d=23400$ s to $d/20$, where $d$
is the daily time horizon at the New York Stock Exchange.
The amount of information processed 
consists of about 100 millions of transactions.
The time horizons investigated 
are $\Delta t=d=6$ h and $30$ min (a trading day time interval), 
$\Delta t=d/2=3$ h and $15$ min, $\Delta t=d/5=1$ h and $18$ min,
$\Delta t=d/10=39$ min and $\Delta t=d/20=19$ min and $30$ sec.
The shortest time horizon was chosen in order to statistically ensure
that for each stock at least $1$ transaction occurs during the time horizon $\Delta t$.
The daily mean number of transactions for the 100 selected stocks
is ranging from 11944.3 transactions of Intel Corp. (INTC)
to the 121.48 transactions of Mallinckrodt Inc. New (MKG).

The `Epps effect' predicts that the intra-sector pair
correlation decreases by decreasing the time horizon $\Delta t$.
In Ref. \cite{Bonanno2001}, authors show that 
the mean correlation coefficient $\langle \rho \rangle$
obtained by averaging over the $n(n-1)/2$ off-diagonal 
elements of the
correlation coefficient matrix
is decreasing when 
$\Delta t$ decreases. The most prominent correlation weakening 
is observed for the most correlated pair of stocks (the ones 
having a correlation coefficient closes to the maximum value 
$\rho_{max}$). In fact, $\rho_{max}$ decreases from 0.76
to 0.52 when $\Delta t$ changes from 6 h and 30 min to 
19 min and 30 s.

The decrease of the correlation between pairs 
of the correlation based network of stocks 
affects the nature of the hierarchical organization of 
stocks. The clusters observed in Fig.~\ref{figA}  progressively
disappear and the arrangement of the minimum
spanning tree moves from a structured and clustered graph 
to a simpler star-like graph. Fig.~\ref{figB} shows the MSTs 
observed at different time horizons ranging from 
$d/20$ to $d/2$. The change of structure of the MST
is indeed dramatic if one considers the role of some 
highly connected
stock such as, in the present case, GE. This stock has a degree,
i.e. a coordination number, equals to $20$
when $\Delta t=d/2=3$ h and 15 min whereas this number grows up to
$61$  
when the time horizon is decreased to $\Delta t=d/20=19$ min and 30 s.

It is worth pointing out that the change in the structure of the
MST and hierarchical tree is not just a simple 
consequence of the 
`Epps effect'. In fact, the changes observed in the structure 
of the MST suggests that the
intrasector correlation  decreases
faster than intersector correlation between pairs of stocks
of the considered portfolio in a intra-day time scale
\cite{Bonanno2001}. 
These results show that the topology of a correlation based 
network can be affected by the sampling time used to monitor
the time evolution of the system. In other words, the 
system presents a non trivial fast dynamics of stock returns
realizing the complex process of the price formation 
occurring in a financial market.

\section{The network of global financial market}
\label{sec:4}

A correlation based network can also be obtained by investigating
index returns of stock exchanges located around the world
\cite{Bonanno2000}.
It is worth pointing out that
the study of the dynamics of stock exchange indices located 
all over the world presents additional difficulties with respect to
the dynamics of a portfolio of stocks traded in 
a single stock market. To cite just two of the most prominent
ones -- (i) stock markets located all over the world have
different opening and closing hours; and (ii) transactions
in different markets are done by using different currencies 
that fluctuates themselves the one with respect to the other.
It is then important to quantify the degree of similarity 
between the dynamics of stock indices of nonsynchronous markets 
trading in different currencies.

Ref. \cite{Bonanno2000} investigates two sets of data -- 
(i) the nonsynchronous time 
evolution of $n=24$ daily stock market indices computed in local 
currencies during the time period from January 1988 to December 1996,
and (ii) the closure value of the 51 Morgan Stanley Capital 
International (MSCI) country indices
daily computed in local currencies or in US dollars in the
time period from January 1996 to December 1999.   
The stock indices used in this research belong to stock markets 
distributed all over the world in five continents.
Here we briefly discuss the results obtained with the
set of Morgan Stanley Capital 
International (MSCI) daily indices computed in local currencies.

An analysis of daily data of closure values recorded
around the world may induce spurious 
correlations introduced just by the different closure times
of different markets. The effects of nonsynchronous
trading in time series analysis are well documented in the 
economic literature \cite{Lo90,Becker90,Lin94}. In fact,
different degrees of correlation between the New York and
Tokyo markets are estimated depending if one consider
the closure - closure between the two markets or the
closure - opening. In particular, it has been empirically
detected that the highest degree of correlation between 
these two markets is observed between the open-closure 
return of the New York stock exchange at day $t$ and 
the opening-closure of the Tokyo stock market at day 
$t+1$ \cite{Becker90}.

\begin{figure}
\resizebox{1.0\columnwidth}{!}{\includegraphics{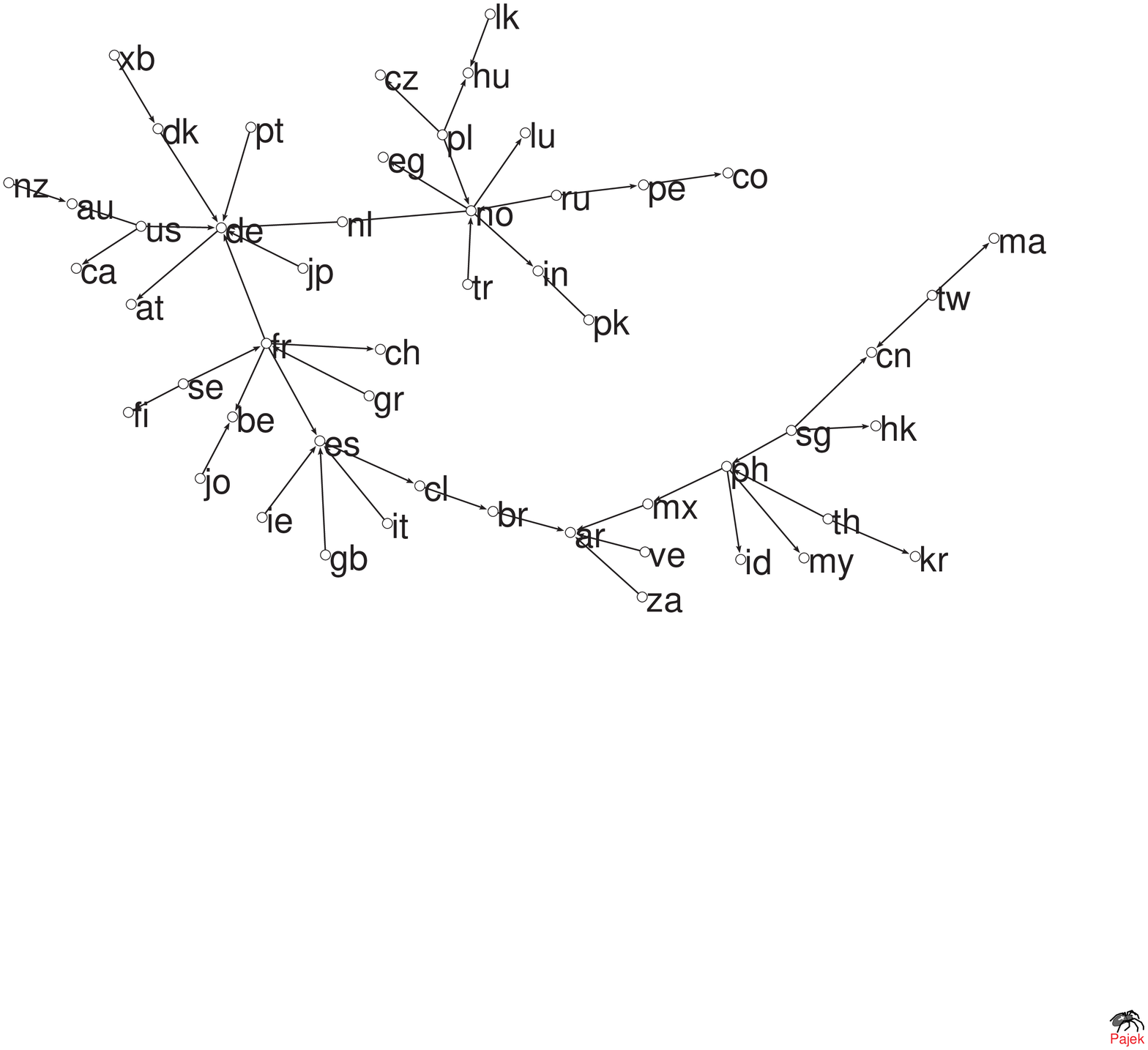}}
\caption{MST of 51 stock exchanges obtained by performing a 
correlation based clustering starting from MSCI index returns
computed in local currencies and by using a time horizon of one week.}
\label{figC}       % Give a unique label
\end{figure}

Ref. \cite{Bonanno2000} overcomes this intrinsic limitations 
by considering a week time horizon so that the nonsynchronous 
hourly mismatch of index data is minimized.
The correlation coefficient is 
computed between all the possible pairs of indices present in 
the database. As usual, the statistical average is a temporal average 
performed on all the trading weeks of the investigated time period.
Authors obtain the $n \times n$ matrix of 
correlation coefficient for weekly logarithm index differences. 
The $51$ indices investigated in Ref. \cite{Bonanno2000} 
belong to $51$ different countries. They 
comprise the so-called emerged and emerging markets. The indices and
their symbols can be found at the web site http://www.mscidata.com. 
The data are daily data and 
covers the period 1996-1999. In Fig.~\ref{figC}
we show the result of the analysis performed in 
Ref. \cite{Bonanno2000}.

The graph of Fig.~\ref{figC} shows a clear regional clustering.
In fact, one can easily note an European cluster
linked to the North American stock exchanges. These last
stock exchanges are linked to Australian and New Zealand 
stock exchanges. The clusters of South-American
and Asian (with the exception of Japan) stock exchanges
are also clearly recognizable.
Once again, the correlation based network shows 
clusters organized with respect to an ordering principle,
which is in this case the regional location of 
stock exchanges.
However, the topological properties of the graph
are pretty different from the one observed
for stock returns of a portfolio traded in a financial
market. In fact, the graph is characterized by a low number
of the average degree of elements. Moreover,
differently from the case of the portfolio of stocks,
the elements characterized by a relatively high coordination number 
do not coincides with the most capitalized stock exchanges.
      
In summary, Ref. \cite{Bonanno2000} has shown that 
sets of stock index time series
located all over the world can provide a correlation based
network that is showing a regional clustering but it
is characterized by topological properties pretty different
than the one observed in a portfolio of stocks traded in
the same financial market.

\section{Networks of volatility time series}
\label{sec:5}

Another investigation has been devoted to detect the 
network of relation which is present  
among volatility time series of stock prices traded in
a financial market. Volatility is
a key financial quantity controlling the risk profile
of a given financial asset traded in a market \cite{Campbell97}. 

In Ref. \cite{Micciche2003} some of us investigate the 
statistical properties of cross-correlation of volatility
time series for the $93$ most capitalized stocks traded
in US equity markets during a $12$ year time period. 
Data cover the whole
period ranging from January $1987$ to April $1999$ 
($3116$ trading days). 
In the cited study daily data are considered. 
In particular, authors use for the
analysis the open, close, high and low price recorded for 
each trading day  for
each considered stock.
Starting from the daily price data, volatility 
$\sigma_i(t)$ is
computed by using the proxy $\sigma_i(t) = 2~[\max\{P_i(t)\} - \min\{P_i(t)\}]
/ [\max\{P_i(t)\} + \min\{P_i(t)\}]$ where $\max\{P_i(t)\}$ and
$\min\{P_i(t)\}$ are respectively the highest and lowest price of the stock $i$
at day $t$. 
It should be noted that there is an essential difference between price return and volatility
probability density functions. In fact, the probability density function of
price return is an approximately symmetrical function whereas the volatility
probability density function is significantly skewed.  Bivariate variables
whose marginals are very different from Gaussian functions can have linear
correlation coefficients which are bounded  in a subinterval of $[-1,1]$
\cite{Embrechts2002}.  Since the empirical probability density function of
volatility is very different from a Gaussian, the use of a robust 
nonparametric correlation coefficient is more appropriate for quantifying
volatility  cross-correlation.  In fact, the volatility MSTs obtained  starting
from a Spearman rank-order correlation coefficient are more stable 
than the ones obtained starting from 
the linear (or Pearson's) correlation coefficient \cite{Micciche2003}.
\begin{figure} 
\begin{center} 
\resizebox{1.0\columnwidth}{!}{\includegraphics{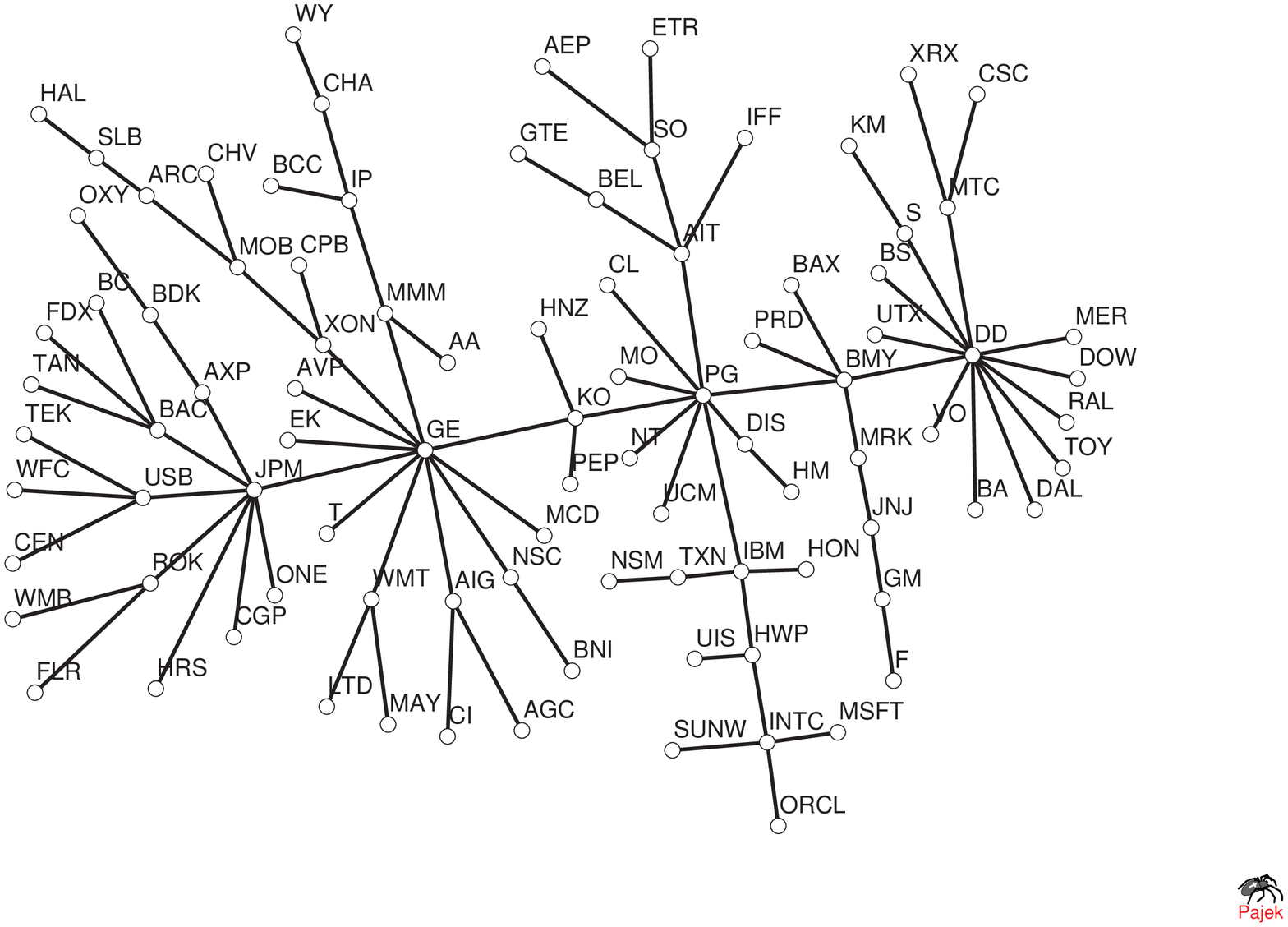}}
\caption{Minimum spanning tree obtained by
considering the volatility time series of 93 mostly capitalized stocks traded
in the US equity markets in August 1998. Each stock is identified by its tick
symbol. The correspondence with the company name can be found in any web site
of financial information. The volatility correlation among stocks has been
evaluated by using the Spearman rank-order correlation coefficient. The MST has
been drawn by using the Pajek package for large network analysis
http://vlado.fmf.uni-lj.si/pub/networks/pajek/ } 
\label{figD} 
\end{center}
\end{figure}   
An example of the MST obtained starting from the volatility time
series and by using the Spearman rank-order correlation coefficient is shown in
Fig.~\ref{figD}. 
A direct inspection of the MST shows the existence of well
characterized clusters. Examples are  the cluster of technology companies (HON,
HWP, IBM, INTC, MSFT, NSM, ORCL, SUNW, TXN and UIS) and the cluster of energy
companies (ARC, CHV, CPB, HAL, MOB, SLB, XON). As already observed in the MST
obtained from the price return time series,  the volatility MST of Fig.
\ref{figD} shows the existence of highly connected stocks. 
Examples are GE, JPM, and DD.
The topology of the network is not too different from the
topology of the network obtained from return time series sampled 
at the same time horizon (Fig~\ref{figA}). Investigations on large sets of 
stocks would be needed to estimate if a quantifiable topological 
difference exists between return and volatility correlation based
networks.

\section{Topology of networks in financial markets}

In the previous sections, we have discussed 
the shape and topology of several networks
obtained by using a correlation based clustering procedure.
In all cases, networks are carrying a clear economic meaning. However
a difference in the topological properties is sometime 
observed when the set of data is ranging from stock portfolios
to a set of stock indices or to the volatility time series 
of a stock portfolio. 
The topological properties are also sensitive to the 
sampling time of the time series used to compute the 
correlation coefficient matrix.
It is therefore worth to investigate 
more deeply the relation between the topological property
of correlation based networks and some simple but widespread
market models.

In Ref. \cite{Bonanno2003} some of us compare the topological properties 
of the MST of empirical data recorded at the New York Stock
Exchange with MSTs obtained from
simple models of the portfolio dynamics. Specifically, authors
consider a model of uncorrelated Gaussian return time 
series and the widespread one-factor model.
This last model is the starting point of the Capital
Asset Pricing Model\cite{Campbell97}.
The topological characterization of the correlation based MST
of real data was originally investigated in Ref. \cite{Vandewalle2001}.
In their study, authors investigated a portfolio of 
approximately 6000 stocks by estimating the correlation 
coefficient on a yearly time period by using approximately 250 
daily data.
Here we discuss the results obtained in 
the study of Ref. \cite{Bonanno2003}, where authors use a smaller 
number of stocks $n$ and a larger number of daily records $T$. 
This choice is motivated by the 
request that the correlation matrix be positive definite.
In fact, when the number of variables is larger than the 
number of time records the covariance
matrix is only positive semi-definite\cite{Mardia79}. 

\begin{figure} 
\begin{center} 
\resizebox{0.75\columnwidth}{!}{\includegraphics{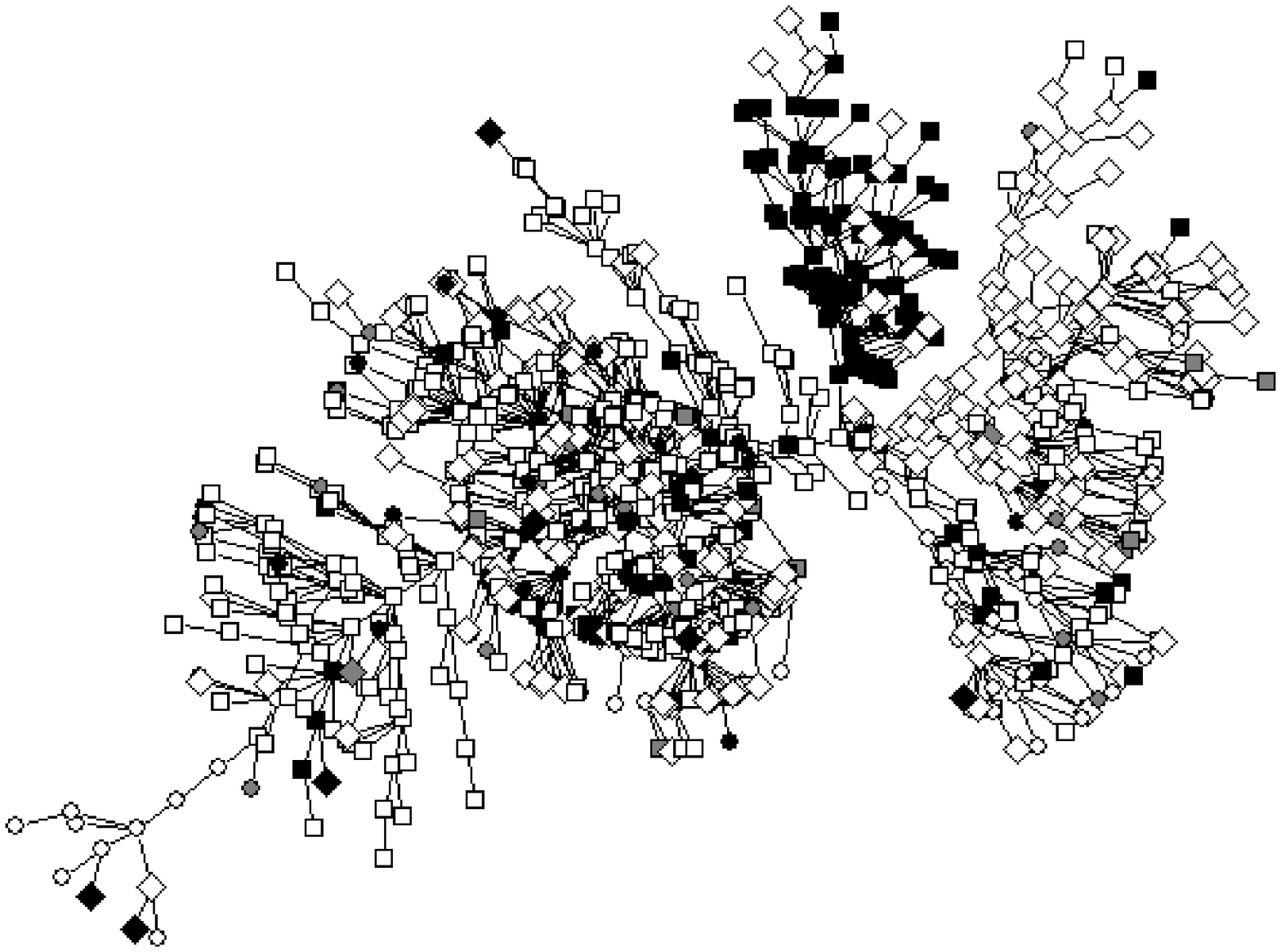}}
\caption{MST of real data from daily stock returns of $1071$ stocks for the 
$12$-year period 1987-1998. The node symbol is based on the Standard
Industrial Classification system. For the correspondence see the text.} 
\label{figE} 
\end{center}
\end{figure}

The data set used in Ref. \cite{Bonanno2003} 
consists of daily closure prices for 1071 stocks 
traded at the NYSE and continuously present in the 12-year period 1987-1998 
(3030 trading days). The ratio $T/N \simeq 2.83$ is
significantly larger than one and the correlation matrix 
is positive definite.
Fig.~\ref{figE} shows the MST of the real data. 
The symbol code is chosen by using the main industry sector of each 
firm according to the Standard Industrial Classification 
system
for the main industry sector of each firm and the correspondence is 
reported in the figure caption.
Again regions corresponding to different sectors are clearly seen
on a very large scale. 
Examples are clusters of companies belonging to the financial sector (white diamonds),
to the transportation, communications, electric gas and 
sanitary services sector (black squares)
and to the mining sector (white circles). The mining sector companies 
are observed to belong
to two subsectors one containing oil companies 
(located on the right side of the figure)
and one containing gold companies (left side of the figure).

\begin{figure} 
\begin{center} 
\resizebox{0.75\columnwidth}{!}{\includegraphics{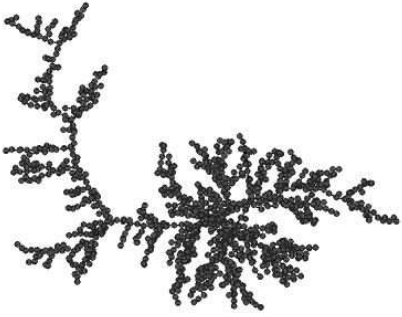}}
\caption{ MST obtained by a realization of a random model of $1071$
Gaussian uncorrelated time series of length $3030$.} 
\label{figF} 
\end{center}
\end{figure}

The empirical MST of real data can be compared with the results 
obtained from simple models of the simultaneous dynamics of
a portfolio of assets. 
The simplest model assumes that the return time 
series are uncorrelated Gaussian time series, i.e. 
$r_i(t)=\epsilon_i(t)$, where $\epsilon_i(t)$ are
Gaussian random variables with zero mean and unit
variance.
This type of model has been considered in 
Ref. \cite{Laloux99,Plerou99} as a null hypothesis in the study of
the spectral properties of the correlation
matrix.  
It is well known both in the financial and in
the econophysics literature that
a random model does not explain the empirical 
observation of financial time series.
This conclusion is consistent with the observation 
that topological properties of MSTs of random
market models are pretty different from the ones
obtained from real data. 
In the MST obtained with the random model few nodes have
a degree larger than few units. In Fig.~\ref{figF} we show one of
this MST obtained for an artificial market described by
a random model.
In Fig.~\ref{figF} it is clear that the MST is 
composed by long files of nodes. These files join at nodes
of connectivity equal to few units (the typical maximal value observed is
close to 7). In other words, a market based on a random model has a network
characterized by a topology essentially different from the 
one observed in real data.

\begin{figure}
\begin{center} 
\resizebox{1.0\columnwidth}{!}{\includegraphics{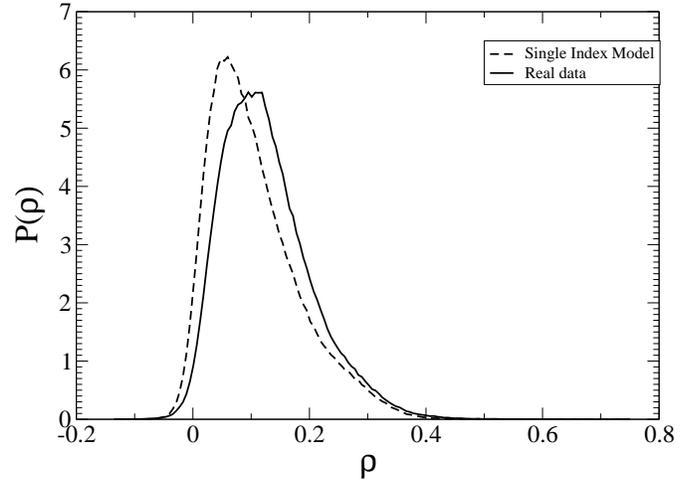}}
\caption{ Empirical probability density function of the correlation coefficients of
a portfolio of $1071$ stocks traded at NYSE in the 12-year period 1987-1998 
(continuous line). The dashed line is the corresponding probability density function 
of a realization of the one-factor model with parameters fitted from real data.} 
\label{figG} 
\end{center}
\end{figure}

A better modeling of the dynamics of a portfolio is 
obtained by using the one-factor model.
The one-factor model assumes that the return of assets 
is controlled by a single factor (or index).
Specifically for any asset $i$ we have
\begin{equation}
r_i(t)=\alpha_i+\beta_i r_M(t)+\epsilon_i(t),
\label{onefact}
\end{equation}
where $r_i(t)$ and $r_M(t)$ are the return of the asset $i$ and of the
market factor at day $t$ respectively, $\alpha_i$ and $\beta_i$ are two
real parameters and $\epsilon_i(t)$ is a zero mean Gaussian noise term
characterized by a variance equal to $\sigma^2_{\epsilon_i}$.
The parameters of the model can be obtained from the real data 
by ordinary least square method. Our choice for the market factor is the
Standard \& Poor's 500 index. The one-factor model is able to reproduce
quite well the distribution of correlation coefficient of the real data.  
In Fig.~\ref{figG} we show the probability density function of correlation
coefficient for real data 
and for the one-factor model.  
It is worth noting that the one-factor model is able to explain 
more that 80\% of the correlation coefficients observed in real data. 
Therefore one could naively expect that also the correlation based MST 
of the one-factor model is quite similar to the correlation based MST of 
the real data.

\begin{figure} 
\begin{center} 
\resizebox{0.75\columnwidth}{!}{\includegraphics{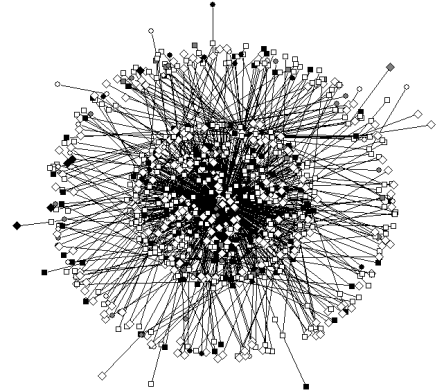}}
\caption{ MST of a numerical simulation of the one-factor model. The symbol 
code is the same as used in Fig.~\ref{figF}.} 
\label{figH} 
\end{center}
\end{figure}

On the contrary the MST obtained with the one-factor model is very different from
the one obtained from real data. 
In Fig.~\ref{figH} we show the MST obtained in a typical realization  
of the one-factor model performed with the control
parameters obtained as described above.
It is evident that the structure of sectors of 
Fig.~\ref{figE} is not present in Fig.~\ref{figH}. 
In fact the MST of the one-factor model has a star-like structure
with a central node. 
The largest fraction of node links directly to the central node and 
a smaller fraction is composed by the next-nearest neighbors.
Very few nodes are found at a distance of three links from 
the central node.   
The central node corresponds to General Electric and the second
most connected node is Coca Cola. It is worth noting that these two stocks
are the two most highly connected nodes in the real MST also.
The reason of the difference between the real and the one-factor model MST (despite
the similarity in the distribution of the correlation coefficients) is attributable
to the noise dressing. A great fraction of the correlation coefficients is heavily
dressed by noise due to the finiteness of the time series. The effect of dressing
is similar in real and in surrogate time series because the length of the time
series has been chosen equal. On the other hand the method used to 
obtain the MST filters part of the relevant information of the correlation 
matrix, discarding the information more heavily dressed by the noise. 
The MST procedure therefore undress the correlation matrix, revealing the
great differences between real and model data.
We want to stress that the difference in the topology between MSTs can be
made more quantitative. In Ref.~\cite{Bonanno2003} some of us conducted 
numerical simulations to show that some topological quantities (the degree
and the in-degree distribution) of real and one-factor MST are different with
$95\%$ statistical confidence.   

In summary, the investigation of the topological properties 
of correlation based networks is able
to discriminate between real data and artificial data
obtained with simple but widespread market models.

\section{Conclusions}

Correlation based networks can be obtained in financial markets
by investigating a certain number of different financial
time series. Here we have reviewed results obtained by us
in different studies. Specifically, the discussed studies 
have been concerning returns of stocks traded in a financial
market at fixed or variable time horizon, volatility time series
and index returns of stock exchanges located all over the world. 
The networks are obtained with a well-defined filtering procedure
\cite{Mantegna99}, which mainly focuses on the most relevant 
correlations among stocks. Different filtering procedures
have been proposed by different authors
\cite{Kullmann2000,Bernaschi2000,Giada2001,Marsili2002} and provide different 
aspects of the information stored in the investigated sets.
The robustness over time of the MST characteristics 
has been investigated in a series of studies 
\cite{Kullmann2002,Onnela2002,Onnela2003,Onnela2003b,Micciche2003}.
The filtering approach based on the MST can also be used to consider
aspects of portfolio optimization \cite{Onnela2003c} and to perform
a correlation based classification of relevant economic entities such
as banks \cite{Marsh2003} and hedge funds \cite{Miceli2003}.

The topology of the correlation based networks depends on the 
investigated set and on the details of investigation 
(an example is the dependence observed for the time
horizon used to compute the stock returns in the investigation
discussed in Sect. 3). The observed topology ranges from the 
star-like one of the top-left panel of Fig.~\ref{figB} to the complex
multi-cluster structure of Fig.~\ref{figA}. Other networks have a
relatively poor number of elements characterized by a high
value of their degree. This last topology may be consistent
with the topology observed in a correlation based network of
a random financial market. On the other hand, the star-like
topology is consistent with a dynamical model defined as
a one-factor model.

In summary, the study of correlation based financial networks
is a fruitful method able to filter out economic information
from the correlation coefficient matrix of a set of
financial time series. The topology of the detected network can be used
to validate or falsify simple, although widespread, market models.

\section{Acknowledgements}

GB, GC, FL, SM and RNM wish to thank partial funding support from 
research projects MIUR 449/97 ``Dinamica di altissima frequenza
nei mercati finanziari", MIUR-FIRB RBNE01CW3M and COSIN IST-2001-33555.

\end{document}